\documentclass[aps,pre,twocolumn,superscriptaddress,showpacs,showkeys ]{revtex4-1}

\usepackage{graphicx}                   
\usepackage{hyperref}                   
\usepackage{amsmath,amssymb}            
\usepackage{epstopdf}
\usepackage{subcaption}					

\usepackage{color}
\definecolor{orange}{rgb}{1,0.5,0}
\definecolor{brown}{rgb}{0.65, 0.16, 0.16}
\definecolor{phlox}{rgb}{0.87, 0.0, 1.0}

\graphicspath{{figs/}}					

\begin{document}

\title{The effect of retardation in the random networks of excitable nodes \\
	embeddable in the Euclidean space}

\author{M. N. Najafi*}
\affiliation{Department of Physics, University of Mohaghegh Ardabili, P.O. Box 179, Ardabil, Iran}
\email{morteza.nattagh@gmail.com}

\author{M. Rahimi}
\affiliation{Department of Physics, University of Mohaghegh Ardabili, P.O. Box 179, Ardabil, Iran}
\email{milad.r.m@chmail.ir }

\begin{abstract}
Some features of random networks with excitable nodes that are embeddable in the Euclidean space are not describable in terms of the conventional integrate and fire model (IFM) alone, and some further details should be involved. In the present paper we consider the effect of the retardation, i.e. the time that is needed for a signal to traverse between two agents. This effect becomes important to discover the differences between e.g. the neural networks with low and fast axon conduct times. We show that the inclusion of the retardation effects makes some important changes in the statistical properties of the system. It considerably suppresses/restricts the amplitude of the possible oscillations in the random network. Additionally, it causes the critical exponents in the critical regime to considerably change. 
\end{abstract}


\maketitle

\section{Introduction}
Complex networks today have a wide applications in science, ranging from neuroscience~\cite{boccaletti2006complex} and intelligent signal processing~\cite{zaknich2003neural}, to the social networks and World Wide Web~\cite{latora2017complex}. In the neural networks, the theories of adoptive optimizing control can be served as a basis for the learning process which, in the behavioral sense, is driven by changes in the expectations about the future salient events such as rewards and punishments~\cite{schultz1997neural}. There are many models to explain the experimental neuronal avalanches~\cite{beggs2003neuronal} which are based on the Hodgkin-Huxley model~\cite{izhikevich2003simple}. The criticality is a key factor in brain, since it improves the learning~\cite{de2010learning}, optimizes the dynamic range~\cite{shew2009neuronal,kinouchi2006optimal,larremore2011predicting,larremore2011effects}, makes information processing efficient~\cite{beggs2008criticality}, and leads to optimal transmission and storage of information~\cite{shew2011information}. \\
In the current state of research on the complex networks with excitable agents, the communications between the interacting agents are supposed to be \textit{instantaneous}~\cite{zaknich2003neural}. \textit{Instantaneous} here means that the conduction time is independent of the length of the connection, i.e. two signals traversing two unequal connections in length have the same travel period. Apparently the \textit{length} should be meaningful here, and the network should have the capability of being embedded in the Euclidean space. Many instantaneous artificial neural networks (which is a type of massively parallel computing architecture based on brain-like information encoding) with a vast range of applications have been invented, like signal processing and also pattern recognition~\cite{zaknich2003neural}. In the systems which are embedded in the Euclidean space, when the speed of the signal (whatever it is) between two agents is very higher than the characteristic speeds in the system (resulting from the speed of the activities of the agents), this approach works well, as can be seen in the partial success of the instantaneous models in describing some experiments on brain~\cite{beggs2003neuronal,levina2007dynamical}, like the self-organized criticality mode of brain activity~\cite{de2006self,hesse2014self}, the chaos for balanced excitatory and inhibitory activity~\cite{van1996chaos}, the neuronal coherence~\cite{fries2005mechanism}, and the synchronization of cortical activities~\cite{singer1993synchronization}. Also theoretical explanation of the neuronal avalanches which are seen in the cerebral cortex (in which the spontaneous neural activities occur at the critical state) are based on such an instantaneous dynamics~\cite{petermann2009spontaneous}. The instantaneous mechanisms which have also been proposed to explain the signal propagation in neocortical neurons based on the repetitions of spontaneous patterns of synaptic inputs should be added to this list~\cite{ikegaya2004synfire}.\\
There are however some situations that this speed is not that high, and one should take the retardation effects into account. Here the various time scales play a vital role. For example, there are many time scales for the neurons in primary auditory cortex of cats, ranging from hundreds of milliseconds to tens of seconds~\cite{ulanovsky2004multiple}. For a general argument on the time scales see~\cite{lemke2000across}. In Ref.~\cite{kiebel2008hierarchy} the hierarchy of time scales in the brain has been considered and analyzed. These time scales are not however necessarily fixed, and in some situations they can be tuned. An example such a tuning of time scale of neuronal activities has been reported in~\cite{boyden2005millisecond}, for which millisecond time scales has been achieved. A tangible example of the importance of the time scales and conduction times (and correspondingly the speed of signal) is the neural systems whose constituents (neurons) are lacking the myelin sheath in which the nerve conduction velocity in the avalanche pulse dynamics are not that high~\cite{kandel1995essentials}. Axon conduction time is definitely a relevant quantity in these networks. This nerve conduction velocity can also be precisely regulated with internal mechanisms, correct exertion of motor skills, sensory integration and cognitive functions~\cite{seidl2014regulation}. In a neuronal population if the conduction velocity is low or equivalently the length of the axons is comparable with the speed of signal times the characteristic times of neuronal activities, the retardation effects become important. The retardation in a nervous system is the effect in which the present activity of neuron in related to the activity of its neighboring neuron at $\delta t=r/v$ times ago in which $r$ is the distance between two neurons and $v$ is the speed of the signal. \\
These retardation effects are expected to be an important in the neural systems with low-speed neurons. For the neuronal cells with the Myelin sheath around their axons (a fatty insulating later that surrounds the nerve cells of jawed vertebrates, or \textit{gnathostomes}) the speed of the signal is higher than that ones without myelin sheath. This causes a lot of differences in these systems, which separates jawed vertebrates from the invertebrates. In vertebrates, the rapid transmission of signals along nerve fibers is made possible by the myelination of axons and the resulting salutatory conduction in between nodes of Ranvier~\cite{seidl2014regulation}.  Among the vertebrates also, the speed of neuronal signals are more or less higher for more intelligent species. Myelination not only maximizes conduction velocity, but also provides a means to systematically regulate conduction times in the nervous system, which to date, has not been understood well~\cite{seidl2014regulation}. Node assembly, internode distance and the diameter of axon, which are controlled by myelination glia, determine the speed of signals along axons. All of these show the importance of the retardation effects in real neural systems.\\
In this paper we consider these retardation effects for a random network with refractory period, i.e. the agents are prevented to send a signal immediately after spiking. Our numerical results show that the retardation effects, not only change the critical behaviors, but also decrease the oscillatory behaviors of the system. By analyzing the branching ration and other statistical tests we show that the point (in terms of largest eigenvalue of the adjacency matrix $\lambda$) at which the critical behaviors starts, the interval of critical behaviors and the point at which the bifurcation begins is just the same as the instantaneous random system. The tuning of the signal propagation in such random networks in therefore promising for controlling some undesirable oscillatory responses.\\
The paper has been organized as follows: In the next section we explain the effects of the retardations and also the method to enter it in the calculations. The SEC.~\ref{NUMDet} has been devoted to the numerical results and the explanation of the behaviors of the model. We close the paper by a conclusion.

\section{Retardation effects}

In this section we consider a random undirected graph with $N$ excitable nodes. Each two nodes are connected with the probability $q$, which results to the average node degree $\left\langle k\right\rangle =qN$. The connections are weighted with quenched random numbers $w_{i,j}$ (between nodes $i$ and $j$ which are connected), whose distribution is uniform in the interval $[0,2\sigma]$, in which $\sigma$ is an external parameter. The state of a node ($i$) at time $t$ is described by $A_i(t)$ which is called the activity, assuming two values: active $A_i=1$ or quiescent $A_i=0$. The aggregate activity at time $t$ is defined as $x(t)\equiv\sum_{i=1}^N A_i(t)$, which is commonly used for analyzing the statistical properties and also the stability of the system. According to the integrate and fire model the $i$th node at time $t$ becomes active depending on the aggregate input signal:
\begin{equation}
p(A_i(t)=1)=f\left( \sum_{j}w_{i,j}A_j(t-1)\right) 
\end{equation}
in which $f$ is a dynamical (monotonically increasing) map which yields the probability that a node becomes active based on the input signal to that node, and is commonly chosen to be:
\begin{equation}
f(y)=\begin{cases}
y, & 0\leq y \leq 1\\
1 & y>1
\end{cases}
\end{equation}
This dynamics is known to be dictated by the largest eigenvalue of the adjacency matrix $w_ij$, namely $\lambda$~\cite{larremore2011effects,larremore2012statistical}. For the random graph that has been considered in this work, this eigenvalue is equal to $\lambda=\sigma\left\langle k\right\rangle =\sigma qN$~\cite{larremore2012statistical,restrepo2007approximating,moosavi2017refractory}. For $\lambda<1$ the system has an attractor $x=0$, i.e. some (stochastic) time after the external local drive, all nodes of the system become inactivate. The completely inverse behavior is seen for $\lambda>1$ in which the perturbation grows with time, reaching $x=N$ at some stochastic time. The intersection between these two intervals, i.e. $\lambda=0$ is known to be critical for which some power-law behaviors occur for e.g. the avalanches ($\equiv$ an overall process between starting and ending an activity). Let us define $S$ as the avalanche size ($\equiv$ the total number of activities in an avalanche), $M$ as the avalanche mass ($\equiv$ the total number of distinct nodes which have been activated (at least once) in an avalanche), $D$ as the avalanche duration ($\equiv$ the total time interval of an avalanche), and $x$ the active nodes at a given time. Then the fingerprint of the criticality can be found in the power-law behavior of the distribution functions, i.e. $N(\zeta)\sim \zeta^{-\tau_{\zeta}}$ in which $\zeta=S,M,D,x$. Also the critical point is detectable in terms of the branching ratio which is defined as the conditional expectation value $b(X)\equiv E\left[\frac{x_{t+1}}{X} |x_t=X \right] $, i.e. the expectation value of $x_{t+1}/X$ conditioned to have $x_t=X$. For the critical system $\lim_{X\rightarrow 0}b(X)=0$, and also $\lim_{X\rightarrow 0}\frac{\text{d}b(X)}{\text{d}X}<0$\cite{moosavi2017refractory}. These two conditions state that when $X\rightarrow 0$, $b(X)$ approaches to zero from the negative values. It also determines the possible fixed points of the model in hand by the condition $b(X^*)=1$.\\

Now let us explain the network with refractory period (which has been done in Ref.~\cite{moosavi2017refractory}) and also the retardation effects. It is shown that the inclusion of the refractory period in the dynamics has some nontrivial effects, like the extension of critical interval, bifurcation, and non-trivial fixed points~\cite{moosavi2017refractory}. The retarded integrate and fire model is defined by the following non-linear dynamical equation:
\begin{equation}
p(A_i(t)=1)=\delta_{A_i(t-1),0}f(\text{sum}(A_i(t)))
\label{main}
\end{equation}
in which $\text{sum}(A_i(t))$ is the integrated effect which has arrived to the $i$th site at time $t$, taking into account the retardation effects. Also $\delta_{A_i(t-1),0}$ is unity if $A_i(t-1)=0$ and is zero otherwise, i.e. it is the effect of the refractory period in the node. The sum-function sums the integrated retarded weighted signals, and is defined as:
\begin{equation}
\text{sum}(A_i(t))\equiv \sum_{t'=0}^t\sum_{j=1}^N w_{i,j}A_j(t')G_{i,j}(t,t')
\end{equation}
in which we have defined $G_{i,j}(t,t')\equiv\delta(t',t-\frac{|i-j|}{v})$ as the retarded Green function, and $|i,j|$ is the distance of $i$th and $j$the nodes. Therefore, this dynamic works for networks that are embedded in the Euclidean space, that is supposed to be two-dimensional in this study. It is notable that it is not the only way to define $G_{i,j}(t,t')$. For example, one can take into account the dissipation of the signal as a function of the length or the time. By inserting this into Eq.~\ref{main}, one finds that:
\begin{equation}
p\left( A(t)=1\right) =\delta_{A_i(t-1),0}f\left[  \sum_{j=1}^Nw_{i,j}A_j\left( t-\frac{|i-j|}{v}\right)\right] 
\label{main2}
\end{equation}
This function carries instantaneously the effects of retardation and the refractory period, and $w_{i,j}$ is a periory known quenched stochastic variable that was introduced above. 

\subsection{Numerical details}

\begin{figure*}
	\centering
	\begin{subfigure}{0.33\textwidth}\includegraphics[width=\textwidth]{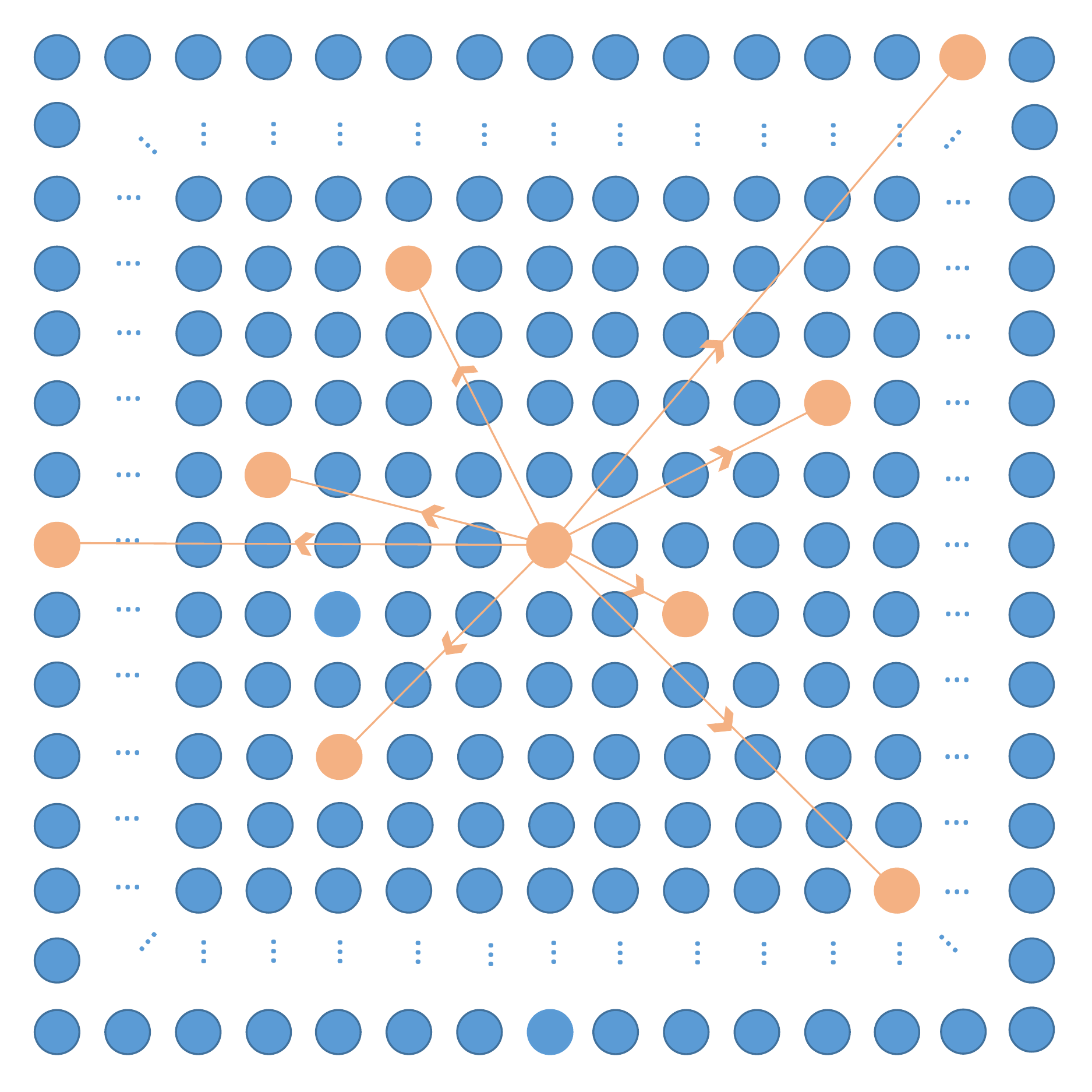}
		\caption{}
		\label{fig:schematic_network}
	\end{subfigure}
	\begin{subfigure}{0.53\textwidth}\includegraphics[width=\textwidth]{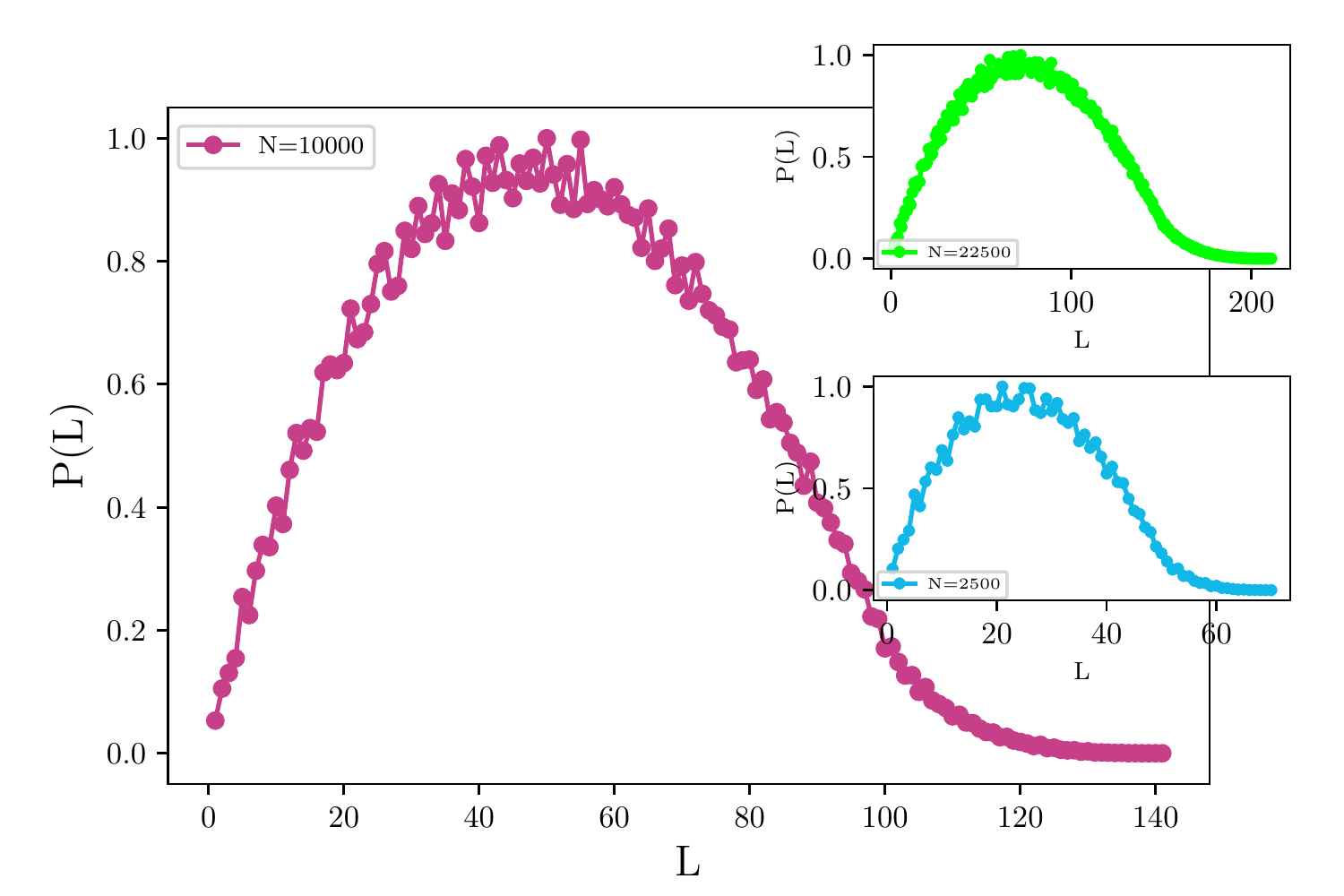}
		\caption{}
		\label{fig:length}
	\end{subfigure}
	\caption{(Color online): (a) The schematic pattern of a highlighted agent in the random network embedded in a two-dimensional space. (b) The distribution of the length between nodes in a simulated random network for $N=100^2$ (main panel), $N=50^2$ (lower inset) and $N=150^2$ (upper inset).}
	\label{fig:definitions}
\end{figure*}

For building the host random network, we simply choose randomly two nodes and connect them. We repeat this $q\frac{N(N-1)}{2}$ times ($\frac{N(N-1)}{2}$ being the total possible links in the system). In Fig.~\ref{fig:schematic_network} we have shown schematically a random graph that has been embedded to two dimensions. The Fig.~\ref{fig:length} shows the histogram of the lengths of the connections ($P(L)$) for $N=50^2, 100^2$ and $150^2$ Erdos-Renyi network (embedded in the Euclidean space) which have their peaks at $L=\frac{1}{2}\sqrt{N}$ as expected. \\

In Ref.~\cite{moosavi2017refractory} it has been shown that the model with refractory period shows three relevant regimes: subcritical regime ($\lambda<1$), extended critical regime $1\leq\lambda\leq2$, and period-2 oscillatory regime $\lambda>2$. We examine the effect of retardation in all of these regimes. We also set the signal velocity to unity, i.e. $v\equiv 1$, since it depends on the scale of the system. In addition to the activity-dependent branching ratio $b(X)$, two kinds of quantities are processed: the distribution function of $\zeta$ and the scaling behaviors $\eta\sim \eta^{\gamma_{\eta \zeta}}$ in which $\eta,\zeta=S,M,D$.\\

\begin{figure}
	\centerline{\includegraphics[scale=.6]{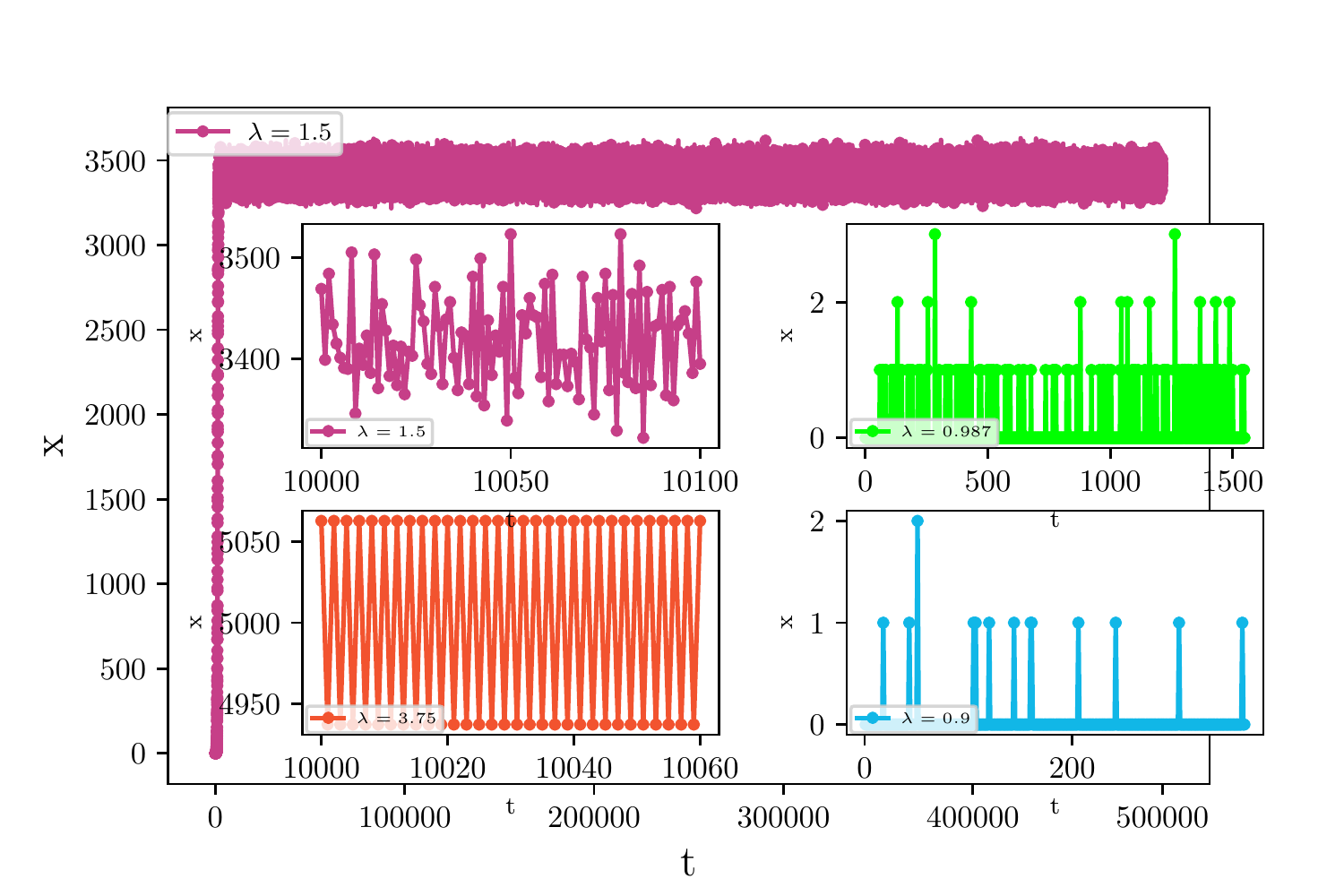}}
	\caption{The time series for the activity (with retardation effects) for various rates of $\lambda$.}
	\label{fig:t-x}
\end{figure}

We should be careful about the definitions. For $\lambda\leq 1$ the stable fixed point is $x^*=0$, whereas for $1<\lambda\leq 2$ $x^*\neq 0$ will be attractor of the dynamics (these fixed points should be obtained by the condition $b(x^*)=1$, and also $\frac{\text{d}b}{\text{d}X}|_{X=x^*}<0$). Therefore, for $\lambda\leq 1$ we have some well-defined avalanches (avalanche $\equiv$ the process in the time interval in which the activity starts from and ends on zero). For $1<\lambda\leq 2$ however we should define the avalanche in another way, since the process does not end, and $x$ fluctuates around $x^*\neq 0$. In this case we define a threshold $X=x^*$ and define the avalanche as the process which starts from and ends on this threshold. The time series for $\lambda=0.9, 0.987, 1.5$ and $3.75$ have been shown as an instance in Fig.~\ref{fig:t-x} for the dynamical system with retardation effects. It is seen that for $\lambda<1$ the stable fixed point is zero, and for $1<\lambda = 1.5<2$ non-zero stable fixed point arises, and also for $\lambda=3.75>2$ the system is in the oscillatory phase. 

All of our analysis In the remaining of the paper will be compared with the results for the system without retardation effects.

\section{measures and results}\label{NUMDet}

\begin{figure*}
	\begin{subfigure}{0.70\textwidth}\includegraphics[width=\textwidth]{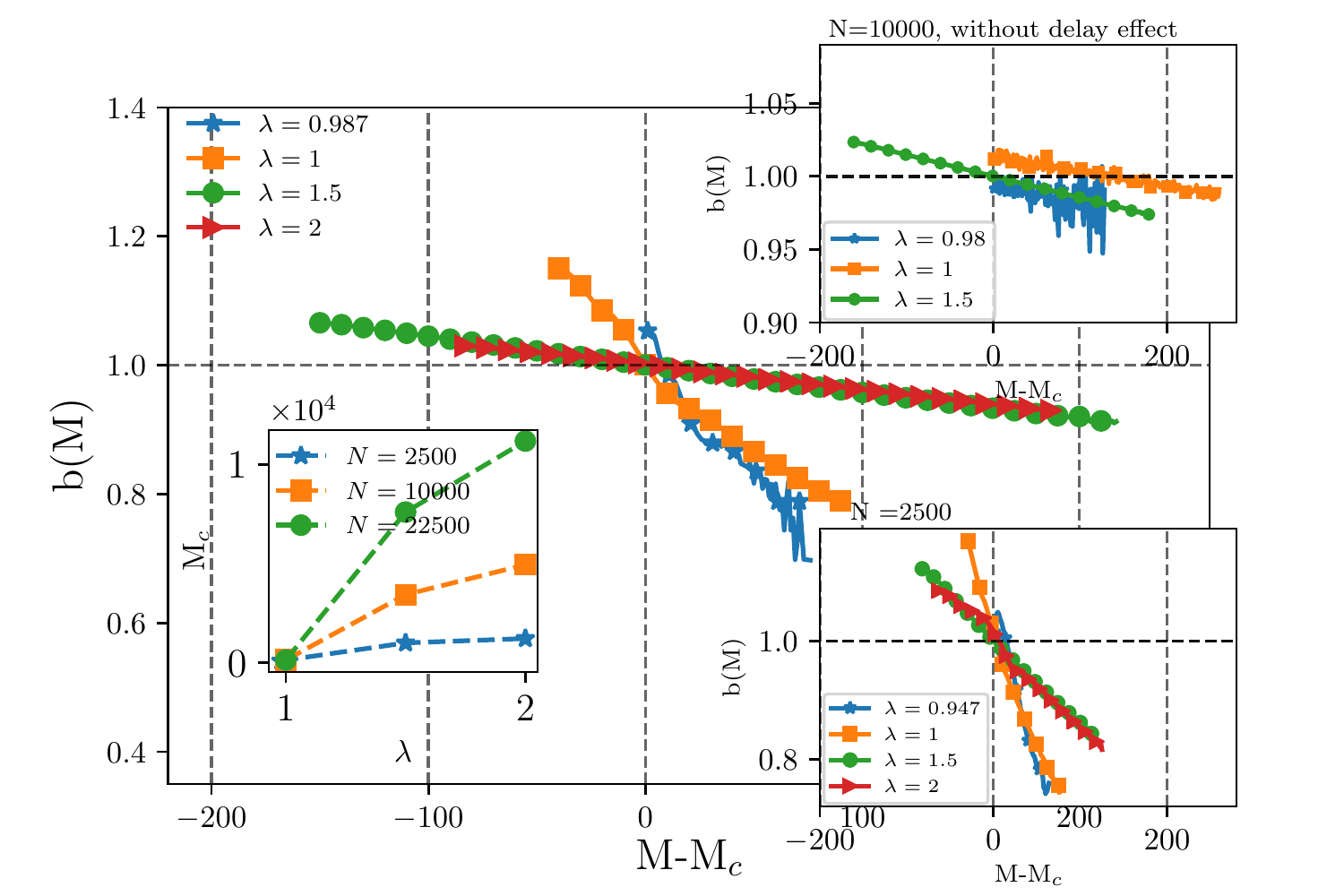}
		\caption{}
		\label{fig:branching_ratio}
	\end{subfigure}
	\centering
	\begin{subfigure}{0.43\textwidth}\includegraphics[width=\textwidth]{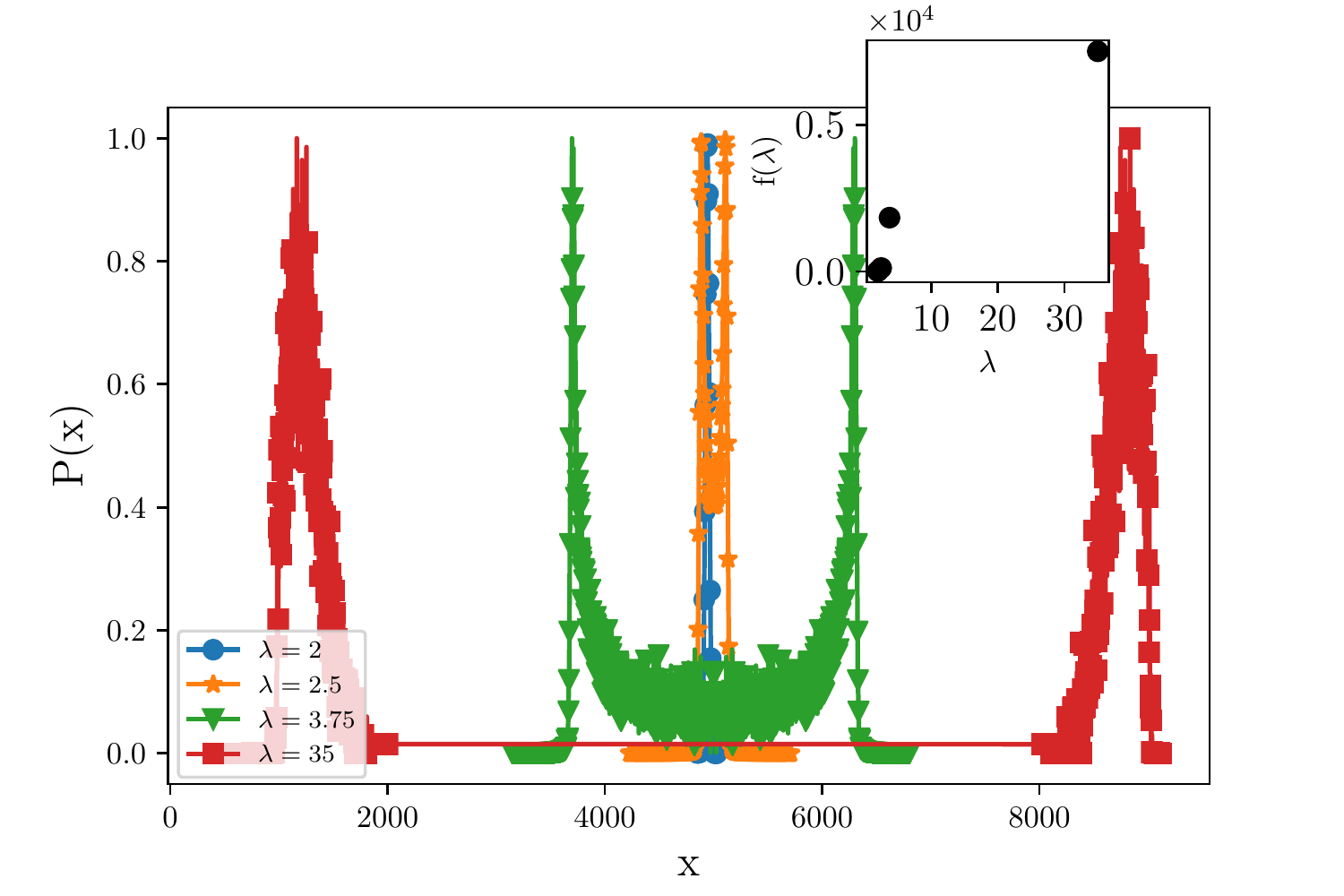}
		\caption{}
		\label{fig:oscillatory_without_delay_effect}
	\end{subfigure}
	\begin{subfigure}{0.43\textwidth}\includegraphics[width=\textwidth]{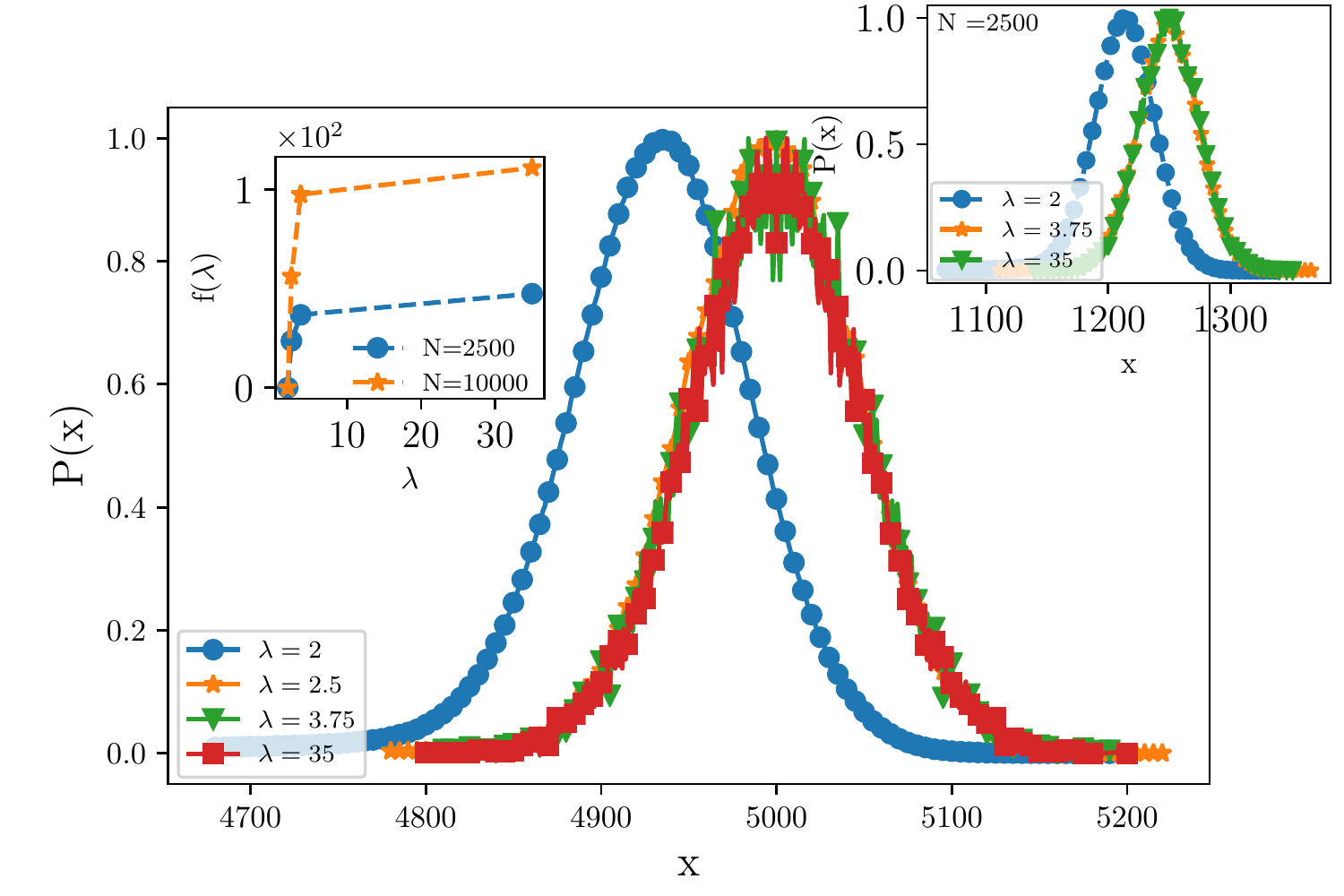}
		\caption{}
		\label{fig:oscillatory_with_delay_effect}
	\end{subfigure}
	\caption{(Color online): (a) The branching ratio $b(M)$ for retarded dynamics (main panel) and instantaneous dynamics (upper inset). Left inset: $M_c$ in terms of $\lambda$ for $N=100^2$. Right inset: The same for $N=50^2$. The distribution function $P(x)$ for retarded dynamics (b) and instantaneous dynamics (c). The inset of (b): $f(\lambda)\equiv \bar{x}_{\text{upper branch}}-\bar{x}_{\text{lower branch}}$ in terms of $\lambda$ for the retarded dynamics.}
	\label{IsingSamples}
\end{figure*}

We first start with the activity-dependent branching ration $b(X)$. In Fig.~\ref{fig:branching_ratio} we have shown this function for both retarded and instantaneous  dynamical systems. We have defined $M_c$ by the relation $b(M_c)=1$. The main panel shows $b(M)$ in terms of $M-M_c$, from which we see that the spoles at the points in which the graphs cross $b(M)=1$ is negative. This confirms that the found points are stable fixed points. In the left inset the $\lambda$ dependence of $M_c$  has been compared for $N=50^2, 100^2$ and $150^2$ networks. It is notable that the mean field results for the instantaneous (retardation-free) dynamical systems reveals that $M_c^{\text{spontaneous}}=N(1-\frac{1}{\lambda})$. The data in this inset is consistent with the mean field result for instantaneous dynamical system, but the (absolute value of) slope of the graphs are very higher, demonstrating that the dynamics (towards the fixed points) is more fast.\\

One of the most serious difference of the retarded and instantaneous dynamical systems arise from their behaviors in the oscillatory regime $\lambda>2$. The Fig.~\ref{fig:oscillatory_without_delay_effect} and Fig.~\ref{fig:oscillatory_with_delay_effect} show the distribution of the activity $x$ for instantaneous and retarded dynamical systems respectively. In the Fig.~\ref{fig:oscillatory_without_delay_effect} the two branches of the oscillations are evident which rapidly grow with $\lambda$. To quantify this, we have plotted $f(\lambda)\equiv \bar{x}_{\text{upper branch}}-\bar{x}_{\text{lower branch}}$, whose numerical value shows the amplitude of the oscillations. $\bar{x}$ is the average value of $x$. We see that it rapidly grows with increasing $\lambda$. The Fig.~\ref{fig:oscillatory_with_delay_effect} however shows a different behavior for the same $\lambda$s. The growth of this amplitude is meaningfully smaller than that for Fig.~\ref{fig:oscillatory_without_delay_effect}. For example, for $\lambda=35$ the gap between two branches is less than $100$, whereas for instantaneous dynamical system, it becomes of order $7000$.\\

Here we see that the inclusion of the retardation effects, restrains the oscillatory behaviors of the random networks with excitable nodes. For neural networks (when is modelled by random excitable nodes) this effect sounds promising for controlling undesirable activity oscillations.\\

\begin{figure*}
	\centering
	\begin{subfigure}{0.49\textwidth}\includegraphics[width=\textwidth]{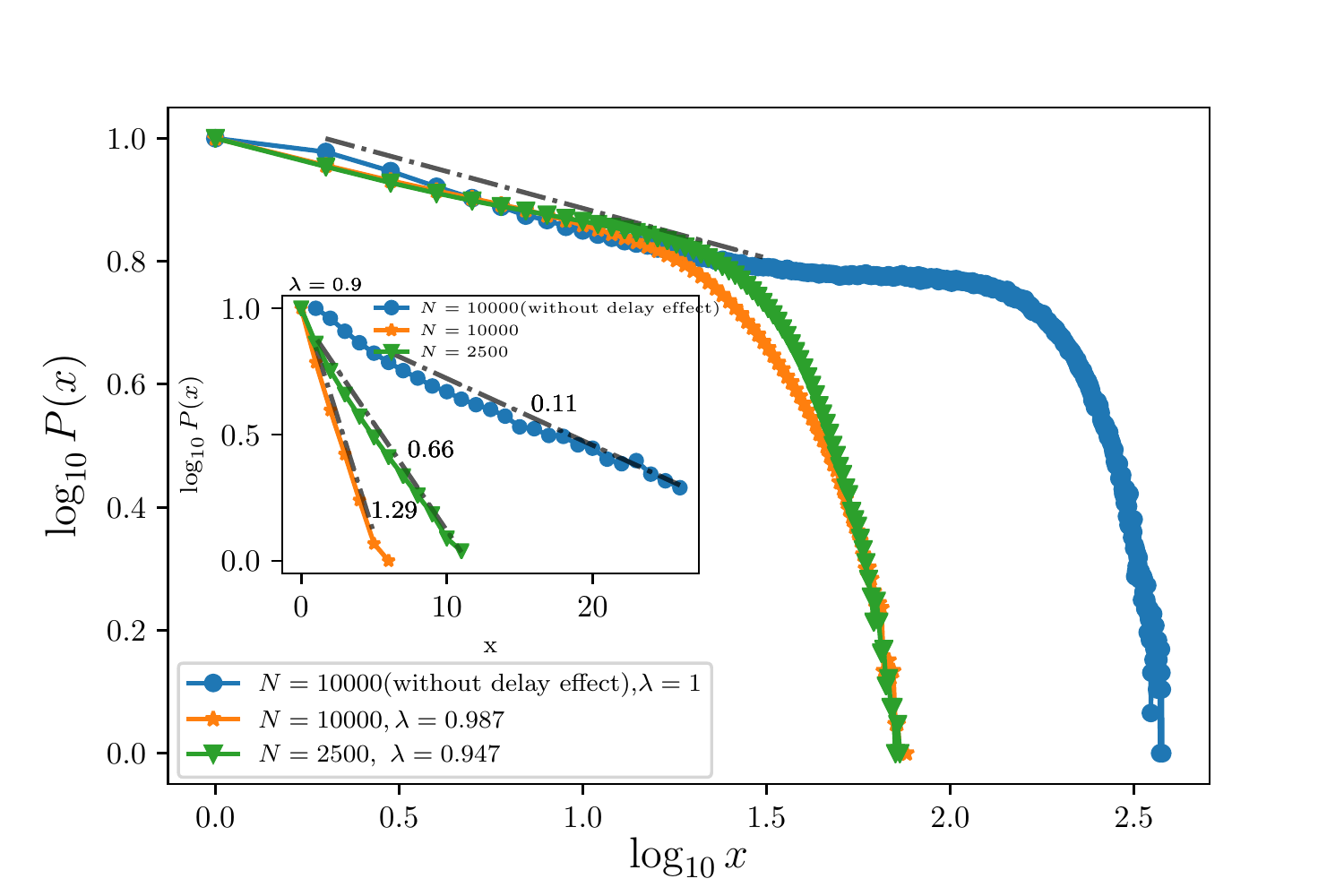}
		\caption{}
		\label{fig:P_x_critical_sub-critical}
	\end{subfigure}
	\begin{subfigure}{0.49\textwidth}\includegraphics[width=\textwidth]{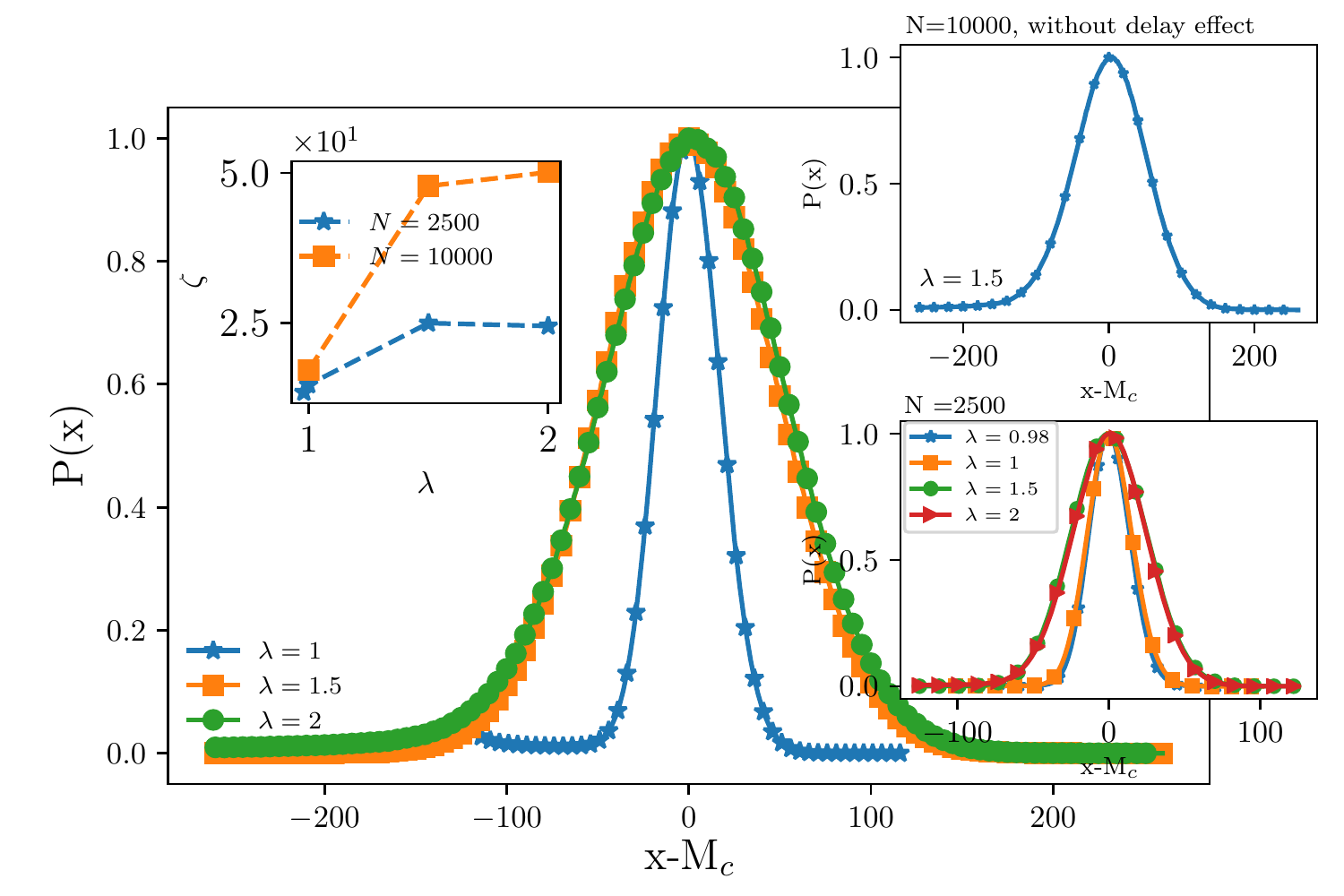}
		\caption{}
		\label{fig:P_x_super_citical}
	\end{subfigure}
	\caption{(Color online): (a) The log-log plot of $P(x)$ for both retarded and instantaneous dynamics for $\lambda$ in the onset of the critical region. Inset: The same for the sub-critical regime. (b) The plot of $P(x)$ for both retarded and instantaneous dynamics for $1<\lambda\leq 2$. Lower inset: the same for $N=50^2$. Upper inset: the same for the instantaneous dynamical system. Left inset: the standard deviation $\zeta$ in terms of $\lambda$. }
	\label{fig:Tc}
\end{figure*}

The same graphs have been shown for the $\lambda$s for the onset of criticality, i.e. in the vicinity of $\lambda=1$ (Fig.~\ref{fig:P_x_critical_sub-critical}) and in the critical interval, i.e. $1<\lambda\leq 2$ (Fig.~\ref{fig:P_x_super_citical}). We see from Fig.~\ref{fig:P_x_critical_sub-critical} that for $\lambda=1$ (or in its vicinity) the exponents of the retarded and instantaneous dynamical systems are nearly the same for the $x$ variable. In the subcritical case however (the inset) $P(x)$ behaves logarithmically for both systems with different (non-universal) slopes. The same has been sketched for  $1<\lambda\leq 2$ in Fig.~\ref{fig:P_x_super_citical}, whose left inset shows that the standard deviations $\zeta$ grows monotonically with $\lambda$. This increase is faster for larger $N$s.\\

\begin{figure*}
	\centering
	\begin{subfigure}{0.49\textwidth}\includegraphics[width=\textwidth]{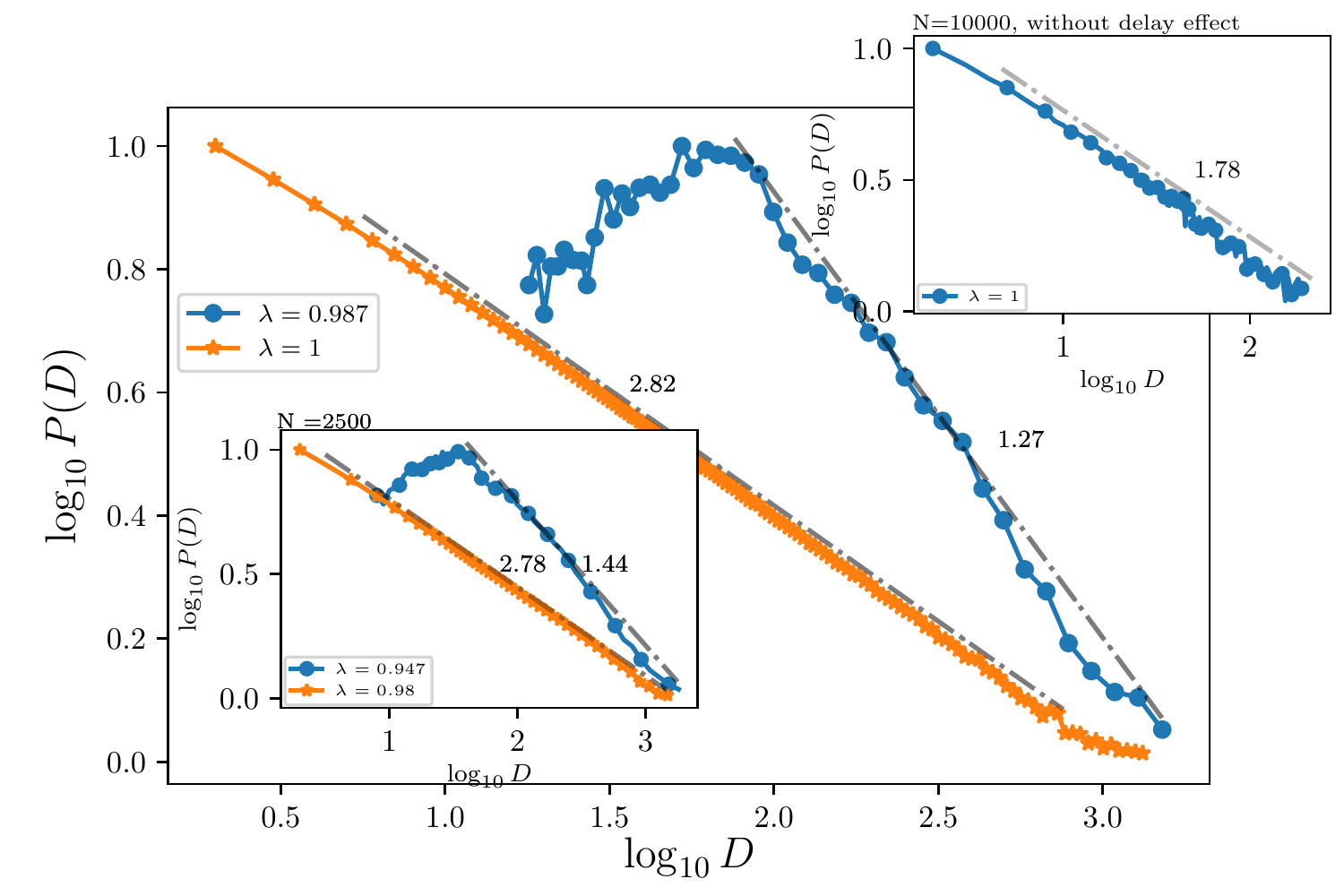}
		\caption{}
		\label{fig:P_D}
	\end{subfigure}
	\begin{subfigure}{0.49\textwidth}\includegraphics[width=\textwidth]{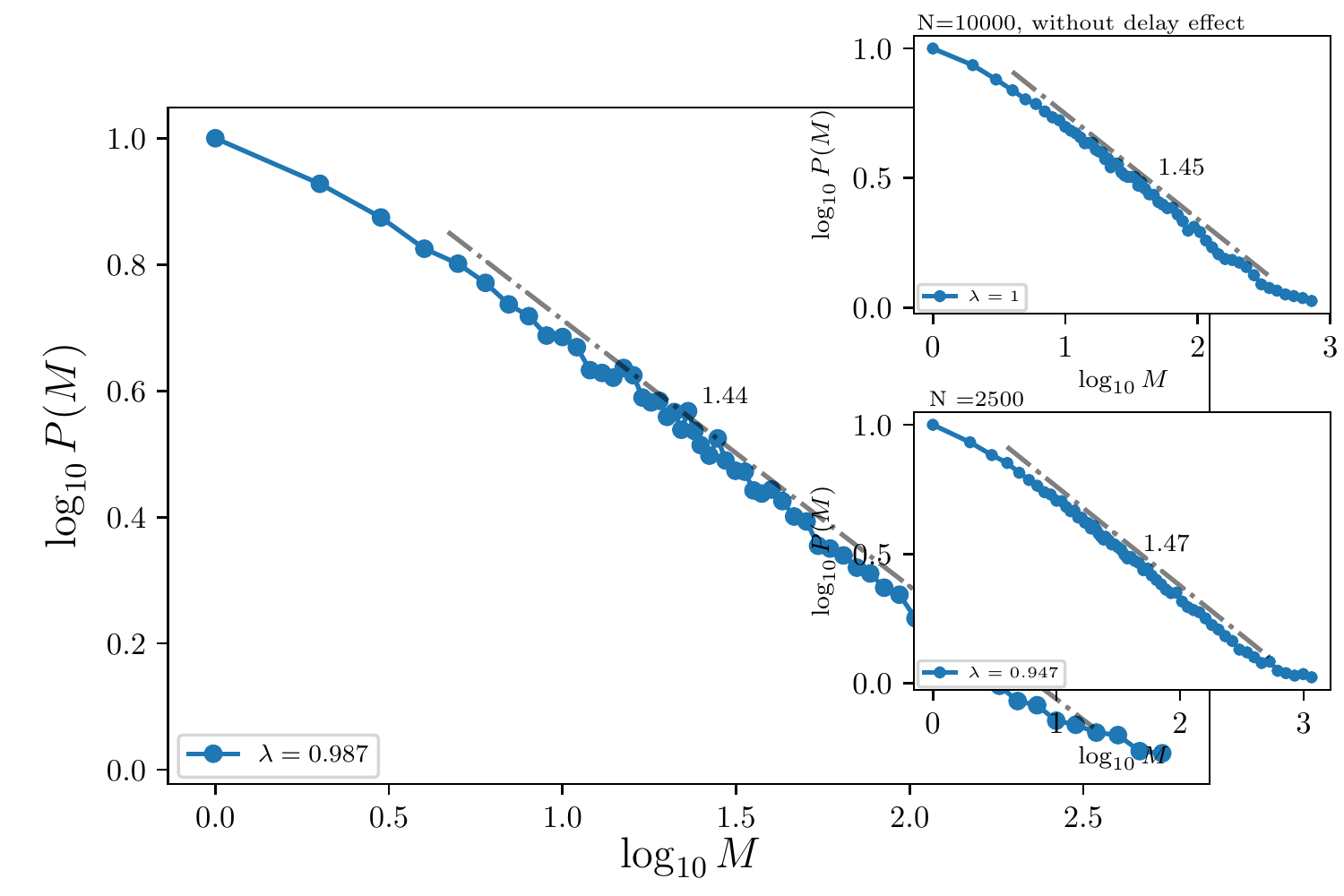}
		\caption{}
		\label{fig:P_M}
	\end{subfigure}
	\begin{subfigure}{0.49\textwidth}\includegraphics[width=\textwidth]{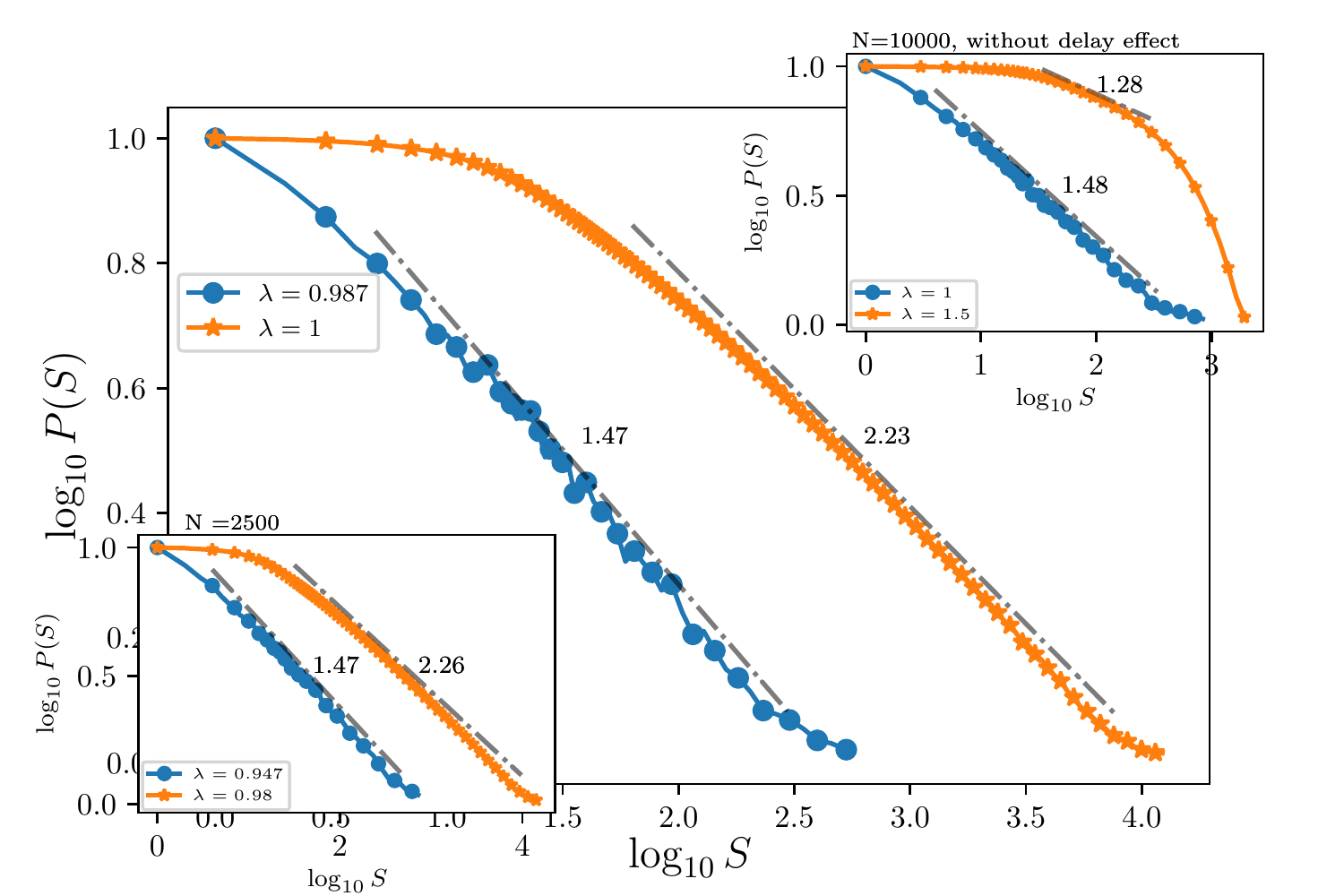}
		\caption{}
		\label{fig:P_S}
	\end{subfigure}
	\begin{subfigure}{0.49\textwidth}\includegraphics[width=\textwidth]{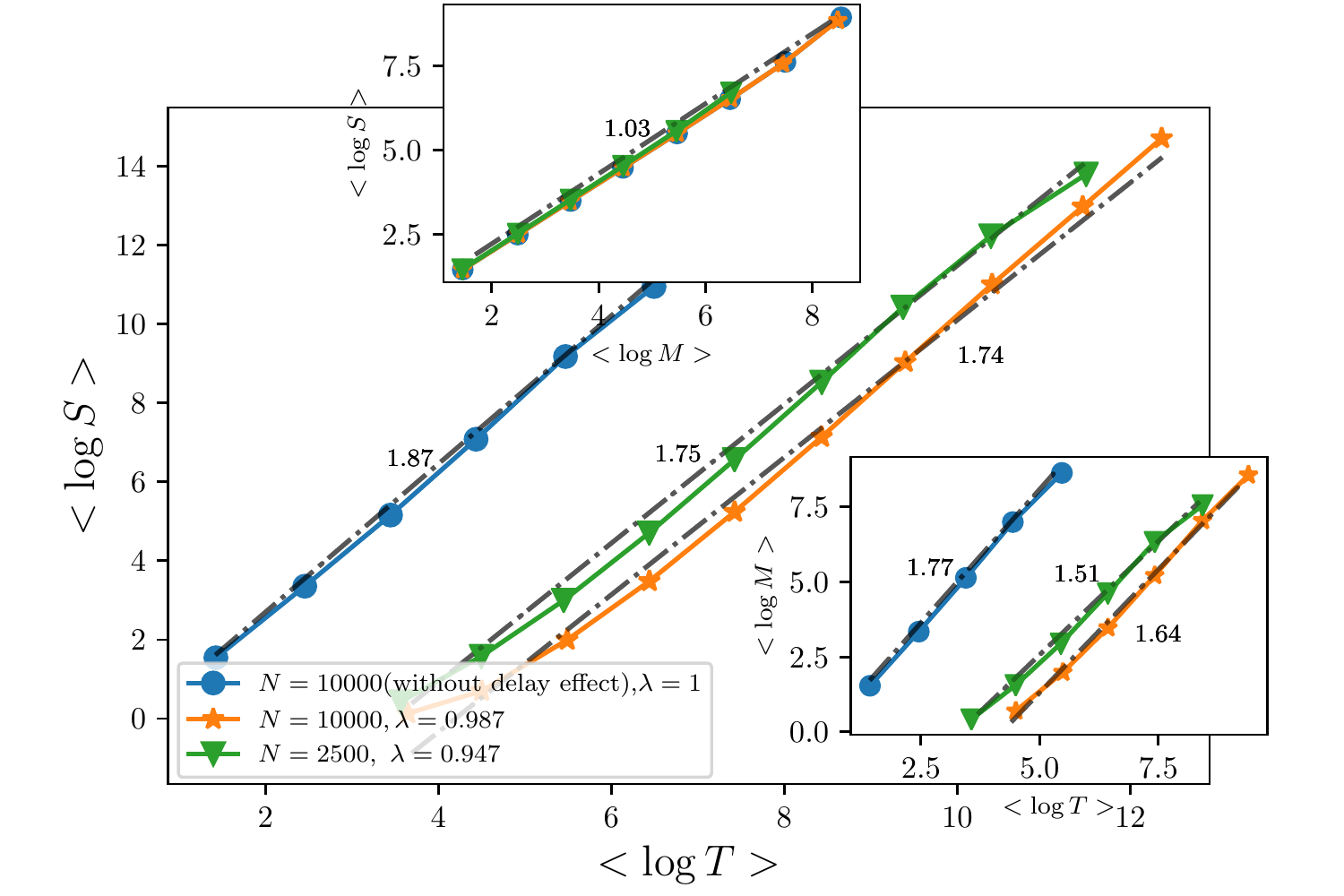}
		\caption{}
		\label{fig:Log_Log}
	\end{subfigure}
	\caption{(Color online) The log-log plot of the distribution functions of (a) the avalanche duration $D$, (b) the avalanche mass $M$, and (c) the avalanche size $S$ for $N=100^2$. Lower insets: the same graph for $N=50^2$. Upper insets: The same for instantaneous dynamical system. (d) The log-log plot of $S-T$ diagram. Upper inset: The log-log plot of $S-M$ diagram.  Lower inset: The log-log plot of $M-T$ diagram.}
	\label{fig:Off-Tc}
\end{figure*}

The retardation is a relevant factor for the statistics of the avalanche duration $D$. More precisely $\tau_D(\lambda=1)$ is very different for retarded and instantaneous dynamics. In the Fig.~\ref{fig:P_D} we see that $\tau_D(\lambda=1)^{\text{retarded}}\approx\tau_D(\lambda=1)^{\text{simulataneous}}+1=2.82\pm 0.1$.\\

The same exponents for $M$ show an agreement with the instantaneous dynamical system for $\lambda$ in the onset of critical region, i.e. $\tau_M(\lambda=1)^{\text{retarded}}\approx\tau_M(\lambda=1)^{\text{simulataneous}}=1.44\pm 0.2$ (Fig.~\ref{fig:P_M}). For the avalanche size $S$ we have $\tau_S(\lambda=1)^{\text{retarded}}=2.23\pm 0.1$ and $\tau_S(\lambda=1)^{\text{instantaneous }}=1.47\pm 0.2$ (Fig.~\ref{fig:P_S}). Note that the determination of the onset of the critical region for a given $N$ has itself an uncertainty and should be determined by analyzing the branching ratio. For example, as the system size $N$ decreases, this value also decreases, e. g. For $N=50^2$ $\lambda^{\text{onset}}= 0.95\pm 0.02$. This itself generates a systematic error in the determination of the exponents on the onset of the criticality.\\

Now let us consider the scaling properties of the statistical variables. This has been done in Fig.~\ref{fig:Log_Log} for all possible scaling quantities. As is explicit in this graph, the scaling between $S$ and $T$ ($T$ being duration of avalanche here) is displaced and the corresponding exponents changes from $1.87\pm 0.05$ (for instantaneous dynamical system) to $1.74\pm 0.05$ (for retarded dynamical system) which is out of its error bar, and the change is meaningful. The same occurs for the $\gamma_{M-T}$ (lower inset), i.e. it changes from $1.77\pm 0.05$ (for instantaneous dynamical system) to $1.64\pm 0.05$ (for retarded dynamical system). Interestingly the $\gamma_{SM}$ does not change considerably and remains on $1.03\pm 0.05$. \\

The critical exponents have been gathered in Table~\ref{tab:exponents} and are compared to the instantaneous dynamical system. 

\begin{table*}
	\begin{tabular}{c | c c c}
		\hline exponent & definition & RDS & IDS \\
		\hline $\tau_D$ & $P(D)\sim D^{-tau_D}$ & $2.82\pm 0.1$ & $1.78\pm 0.1$ \\
		\hline $\tau_M$ & $P(M)\sim M^{-tau_M}$ & $1.44\pm 0.2$ & $1.45\pm 0.2$ \\
		\hline $\tau_S$ & $P(S)\sim S^{-tau_S}$ & $2.23\pm 0.1$ & $1.48\pm 0.2$ \\
		\hline $\gamma_{ST}$ & $S\sim T^{\gamma_{ST}}$ & $1.74\pm 0.05$ & $1.87\pm 0.05$ \\
		\hline $\gamma_{SM}$ & $S\sim M^{\gamma_{SM}}$ & $1.03\pm 0.05$ & $1.03\pm 0.05$ \\
		\hline $\gamma_{MT}$ & $M\sim T^{\gamma_{MT}}$ & $1.64\pm 0.05$ & $1.77\pm 0.05$ \\
		\hline
	\end{tabular}
	\caption{The critical exponents in the onset of criticality of two models: retarded dynamical system (RDS) and instantaneous dynamical system (IDS) for $N=100^2$.}
	\label{tab:exponents}
\end{table*}

\section{Conclusion}
In this paper we have addressed the problem of the effect of retardation in random networks with excitable nodes. The retardation effects have been brought into the calculations using the Eq.~\ref{main2}, which is mixed by the refractory period. We analyzed the branching ratio which yields the possible intervals of distinct behaviors, like the sub-critical, critical and oscillatory behaviors. Our calculations demonstrated that the oscillations are remarkably suppressed by the retardation. This can be a promising effect in the systems that such oscillations are undesirable. \\

Also the critical exponents of the retarded dynamical systems are meaningfully different from that for the instantaneous dynamical systems. The numerical amounts of these exponents can be found in Table~\ref{tab:exponents}.

\bibliography{refs}

\end{document}